\documentclass{ijcaArticle}
\usepackage{amsmath} 
\usepackage{algorithm}
\usepackage{algorithmic}
\usepackage{flushend}
\ijcaVolume{186}
\ijcaNumber{33}
\ijcaYear{2024}
\ijcaMonth{July}

\ijcaVolume{186}
\ijcaNumber{33}
\ijcaYear{2024}
\ijcaMonth{July}
\begin{document}

\title{A Novel Approach to Translate Structural Aggregation Queries to MapReduce Code} 

\author{ 
   \large Ahmed M. Abdelmoniem \\[-3pt]
   \normalsize Department of Computer Science \\[-3pt]
   \normalsize Faculty of Computers and Information \\[-3pt]
   \normalsize Assiut University \\[-3pt]
   \normalsize ahmedcs@aun.edu.eg\\
   \and
   \large Sameh Abdulah \\[-3pt]
   \normalsize Department of Computer Science \\[-3pt]
   \normalsize Faculty of Computers and Information \\[-3pt]
   \normalsize Menoufia University \\[-3pt]
   \normalsize sameh.abdulah@menofia.edu.eg \\[-3pt]
  \and
   \large Walid Atwa \\[-3pt]
   \normalsize Department of Computer Science \\[-3pt]
   \normalsize Faculty of Computers and Information \\[-3pt]
   \normalsize Menoufia University \\[-3pt]
   \normalsize walid.atwa@menofia.edu.eg \\[-3pt]
}

\terms{Big data, MapReduce, Database Managment Systems (DBMS)}
\maketitle

\begin{abstract} 
Data management applications are growing and require more attention, especially in the ``big data" era. Thus, supporting such applications with novel and efficient algorithms that achieve higher performance is critical. Array database management systems are one way to support these applications by dealing with data represented in n-dimensional data structures. For instance, software like SciDB and RasDaMan can be powerful tools to achieve the required performance on large-scale problems with multidimensional data. Like their relational counterparts, these management systems support specific array query languages as the user interface.
As a popular programming model, MapReduce allows large-scale data analysis, facilitates query processing, and is used as a DB engine. Nevertheless, one major obstacle is the low productivity of developing MapReduce applications. Unlike high-level declarative languages such as SQL, MapReduce jobs are written in a low-level descriptive language, often requiring massive programming efforts and complicated debugging processes.
This work presents a system that supports translating array queries expressed in the Array Query Language (AQL) in SciDB into MapReduce jobs. We focus on translating some unique structural aggregations, including circular, grid, hierarchical, and sliding aggregations. Unlike traditional aggregations in relational DBs, these structural aggregations are designed explicitly for array manipulation. Thus, our work can be considered an array-view counterpart of existing SQL to MapReduce translators like HiveQL and YSmart. Our translator supports structural aggregations over arrays to meet various array manipulations. The translator can also help user-defined aggregation functions with minimal user effort.
We also show that our translator can generate optimized MapReduce code, which performs better than the short handwritten code by up to $10.84$X.
\end{abstract}

\section{Introduction}

Large arrays with huge sizes and dimensionalities, such as those used for digital images, simulation data, and statistical results in various fields, including climate/weather, earth science, and geosciences applications, are more suitable for representing big data coming today from numerous sources~\cite{feng2018scidp}. The vast increase in the data volumes of these applications requires more innovative algorithms to manage, query effectively, and process such big data. Array-based data management systems (DBMS) are popular tools for processing a wide range of data that can be represented in multidimensional structures. Examples of existing array-based DBMS include Oracle GeoRaster~\cite{georaster-soft}, MonetDB/SciQL~\cite{idreos2012monetdb}, PostGIS~\cite{hsu2011postgis}, rasdaman~\cite{baumann1998multidimensional}, and SciDB~\cite{brown2010overview}. The primary challenge of using such array DBMS is the intrinsic mismatch between the relational tables representation and the array model~\cite{VLDB97, GRAY05, VLDB05, VLDB08, CIDR09, Brown10, TheFourthParadigm}. Thus, the relational models cannot efficiently represent the array models. 

Recently, array query languages~\cite{AML, AQuery, baumann2021array} and array DBMS provide a solution to bridge the gap between the relational and array models. However, the essential task of these systems is to deal with arrays. Therefore, all field-specific operations in scientific applications should be defined as array operations. This method reduces the complexity of those operations and usually leads to better performance. Herein, the functionality is associated with a set of operations named "structural aggregation," where the array elements are aggregated with a specific group, and the groups are defined based on the positional relationships of the elements. The aggregation matches the structure of the relational models.

On the other hand, MapReduce has served as a popular programming model for large-scale data analysis mainly because of its neat programmability, built-in fault tolerance, and excellent scalability. It has also been leveraged to facilitate query processing and used as a database engine~\cite{gates2009building, lee2011ysmart, wijayanto2020lshape, lee2021knn}. However, one major obstacle around MapReduce is the low productivity of developing entire applications. Unlike the high-level declarative language such as SQL, MapReduce jobs are written in a low-level descriptive language, often requiring massive programming efforts and leading to considerable difficulty in programming debugging~\cite{yang2007map}.

Several translators have been developed to bridge the gap between high-level SQL-like query languages and low-level MapReduce code. These translators aim to automatically generate MapReduce jobs with less programming effort from the developer~\cite{stonebraker2010mapreduce}. Some examples of those translators include Pig Latin~\cite{olston2008pig}, SCOPE~\cite{chaiken2008scope}, HiveQL~\cite{thusoo2009hive} and YSmart~\cite{lee2011ysmart}. In~\cite{lee2011ysmart}, Lee et al. pointed out that SQL to MapReduce translators even contribute more to MapReduce-based query processing than hand-coded programs. For example, the majority of \texttt{Hadoop} jobs on Facebook are generated by Hive automatically rather than developed from scratch manually. Nevertheless, even though array DBMS, as well as an array query language, have recently emerged to be a hot topic in the database community, to the best of our knowledge, so far, no translator has been developed for supporting any high-level array query language used in array databases, which is essentially a counterpart of SQL in relational databases. Therefore, in this work, we propose a novel translator that can translate a SciDB-specialized array query language, Array Query Language (AQL), into MapReduce jobs. Hence, we can gain similar benefits from existing SQL to MapReduce translators like HiveQL and YSmart in a different array database domain. Specifically, we focus on efficiently translating several structural aggregation queries into MapReduce jobs. Specifically, we focus on grid, sliding, hierarchical and circular aggregations. We support both conventional aggregates and predefined aggregates.

In this paper, we propose a novel system that supports translating array queries expressed by \texttt{AQL} in SciDB into MapReduce jobs. We mainly focus on effectively translating several unique structural aggregations. Unlike the traditional aggregations in relational databases, these structural aggregations are designed explicitly for multidimensional array manipulation. Thus, our work can be considered an array-view counterpart of some existing SQL to MapReduce translators like HiveQL and YSmart. Our translator can effectively support array subsetting and aggregation to meet various array queries. Moreover, our translator can support traditional aggregates including {\sc SUM()}, {\sc AVG()}, \\ {\sc COUNT()}, {\sc MIN()}, and {\sc MAX()}. Moreover, with minimum effort, we support predefined aggregation functions like {\sc standard deviation} and {\sc geometric mean}. We also show that our translator can generate optimized MapReduce code, leading to significantly better performance than short hand-written code.

The remainder of this paper is structured as follows: Section 2 gives a brief background on structural aggregations and grouping in multidimensional arrays; Section 3 provides a detailed description of the proposed translation system; Section 4 shows how to optimize the aggregation process using the proposed translator; Section 5 reports different results from our experiments; in Section 6, we summarize the related work and conclude in Section 7.

\section{Structural Aggregations/Grouping}
\label{sec:documentclass}
Standard SQL provides the GROUP BY clause to group operations over a set of transactions by grouping rows with similar values into summary rows. The most common way is to select the candidate set of transactions through value-based grouping, where these transactions share the same value for a specific attribute. This grouping clause can be associated with one or more aggregation functions, for instance, MIN(), MAX(), SUM(), AVG(), and COUNT(). 

A more advanced grouping strategy in SQL is structural grouping, which is performed based on positional relationships. This grouping strategy is only applicable in the case of multidimensional arrays, where this grouping is widely used in scientific applications. For instance, assume calculating the average over elements of a given array. This requires a position-based grouping and an average operation over adjacent elements. In this work, we use the term "structural aggregation" to represent the structural or positional grouping defined over a set of operations. We also enable user-defined functions such as {\sc standard deviation} and {\sc mean} that are widely used in scientific applications.

In this work, we consider three different types of structural aggregations: grid, sliding, and hierarchical (or circular) aggregations. In all of them, we process a section of a given array size (i.e., array slab). The upcoming subsections show different structural aggregations with examples illustrating how aggregation can be applied directly to the array data. The following subsections overview other structural aggregation operations in various scientific applications.

\subsection{Grid Aggregation}
\begin{figure}[h!]
    \centering
    \includegraphics[width=0.6\linewidth]{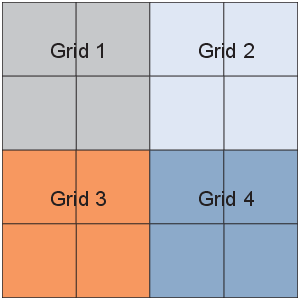}
    \caption{Grid Aggregation (GA)}
    \label{GridAggExamples}
\end{figure}

The primary aggregation used in scientific applications is grid aggregation. Assuming data is represented in a multidimensional array, grid aggregation involves splitting the data array into smaller disjoint grids/blocks. Then, the aggregation operation is performed separately over the values inside each block. For instance, considering astrophysics applications, many astronomy events can be stored in substantial multidimensional arrays. The elements of these arrays represent disjoint spatial information. In this case, the blocks represent spatial grids, and the events histograms can be generated over each grid using aggregation~\cite{EDBT11}. Figure~\ref{GridAggExamples} shows an example of a $4\times4$ array split into four blocks (i.e., a $2\times2$ grid) array elements that can be aggregated for each grid.

\subsection{Sliding Aggregation}
\begin{figure}[h!]
    \centering
    \includegraphics[width=0.7\linewidth]{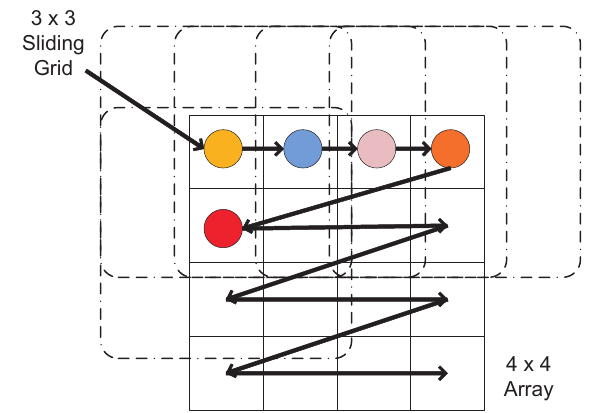}
    \caption{Sliding Aggregation (SA)}
    \label{slidingAggExamples2}
\end{figure}

Sliding aggregation defines a fixed-size grid to the given array, and the target elements are aggregated in a stride way. The stride value is a predefined value with a default value equal to 1. Sliding aggregation is commonly used in earth science applications to perform various operations. For example, image noise removal operations for satellite images using standard algorithms like the non-local means algorithm (ML-means)~\cite{NLMeans}. In this algorithm, the Gaussian core kernel is applied to the chosen sliding grid to smooth the outliers and minimize the impact of the internal noise. Figure~\ref{slidingAggExamples2} shows an example of the sliding aggregation where the predefined grid is a $4\times4$ array where elements in each array slab are aggregated. Crime classification~\cite{bogomolov2015moves} and credit card fraud detection~\cite{jiang2018credit} are two applications that rely on sliding aggregation.

\subsection{Hierarchical - Circular Aggregation}
\begin{figure}[h!]
    \centering
    \includegraphics[width=0.6\linewidth]{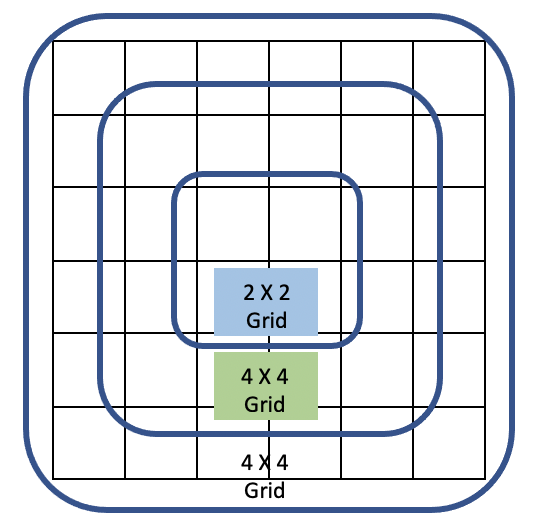}
    \caption{Hierarchical Aggregation (HA)}
    \label{structuralAggExamples3}
\end{figure}

In hierarchical aggregations, the user defines the grid centroids to match the centroid of the entire data array, where a predefined radius is used in the innermost grid and is propagated for other grids. The radius of the innermost grid becomes the base of the fixed step increase, which defines the radius of the outer grids until they reach the array boundaries. Figure~\ref{structuralAggExamples3} gives an example of three grids in orange, blue, and black colors. The aggregation is performed on a different number of elements for each grid. Space science applications are a typical example that relies on hierarchical aggregations. The applied operation may require looking into the gradual effect of radiation sources on specific locations that can lead to an explosion. Hierarchical aggregation over the given data can show the boost in radiation in various regions and forecast potential explosions before occurring.

\begin{figure}[h!]
    \centering
    \includegraphics[width=0.6\linewidth]{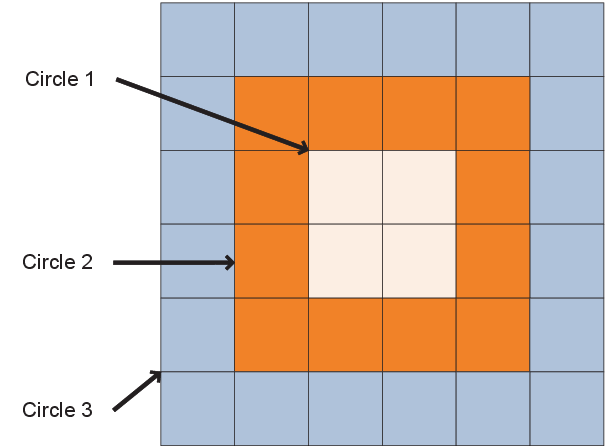}
    \caption{Circular Aggregation (CA)}
    \label{structuralAggExamples4}
\end{figure}

Circular aggregation is hierarchical aggregation that is aggregated in disjoint circles instead of square grids. Figure~\ref{structuralAggExamples4} shows an example of circular aggregation for the example shown in Figure~\ref{structuralAggExamples3}.

\section{AQL to MapReduce Translator}
\label{sys_design}

This section gives a detailed overview of the AQL to MapReduce translator design and implementation, clearly describing the API that supports our predefined aggregation functions.

\subsection{System Overview}

The primary input to our proposed translator is a structural aggregation query and an associated metadata file in AQL format. The translator can parse the query and generate the required MapReduce job script output.

\begin{figure*}[h]
\centering
    \includegraphics[width=0.8\linewidth]{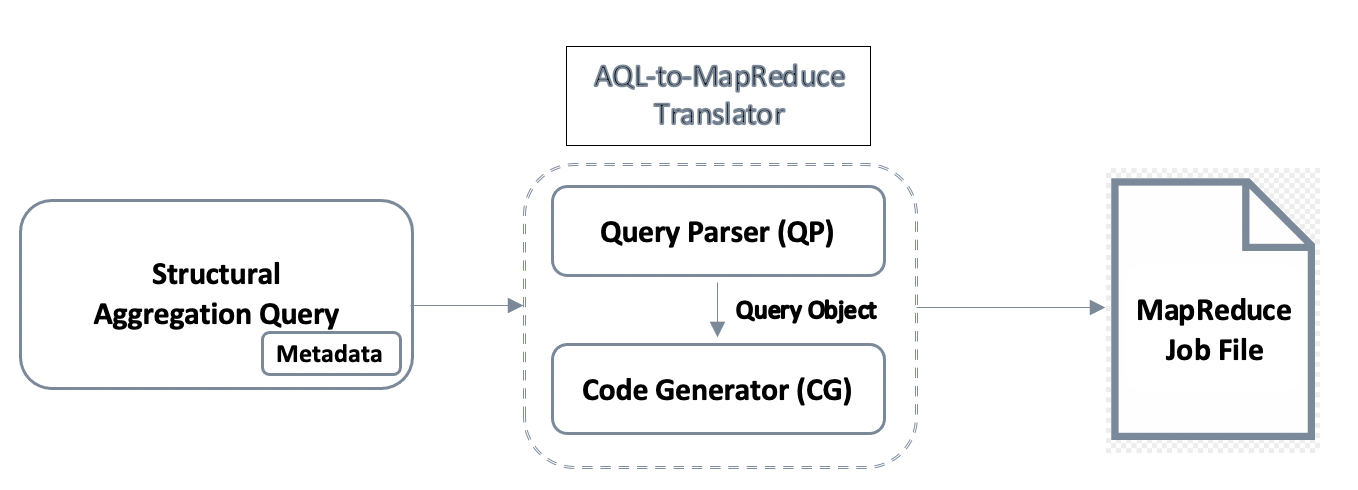}
    \caption{The AQL to MapReduce Translator Components.}
    \label{overview}
\end{figure*} 

Figure~\ref{overview} shows our proposed translator and its execution flow. The execution flow includes two main steps: query parsing, to parse the input query and its metadata file and generate a query object; and code generation, to generate the MapReduce code script with the query object and predefined built-in MapReduce script templates. The input is a structural aggregation query given in AQL format. The translator relies on a query parser to parse the given query as well as the metadata. The parser generates some information, including the array name, the array subset, the aggregation type, and the aggregate operator. In the meantime, with the given array name, the associated metadata file, which is stored in the file system along with array data, is also parsed to retrieve information about the input array, including the array element data type, dimension, and layout. During this parsing, some semantic checks are also performed, e.g., the input array must exist in the system, and the array indices specified by the query must be within the physical layout of the array. The output of this parsing step is a query object that encapsulates all the information needed for processing, including the aggregation operator, array name, dimensionality, dimension indices, value-based predicate, etc.

Afterward, the query object is given to the code generator to generate the MapReduce code. In this step, all the query information provided by the query object is transformed into a query parameter configuration file, which includes all the parameters used by some built-in MapReduce script templates. The fundamental way of the translator is that, for each aggregation type, there exists a particular pattern for composing the MapReduce job, which can be abstracted by a MapReduce script template and pre-stored in the translator. Thus, the built-in templates can limit all the changes of MapReduce code for different aggregation queries. Finally, the generated parameter configuration file and the MapReduce script template are taken as the output, i.e., the MapReduce job.

\subsection{Supporting User-Defined Aggregation Function API}
Aggregation in multidimensional arrays in scientific applications is always domain-specific and can be complex. Thus, defining aggregation functions in advance is essential. In our proposed translator, the user can predefine a set of aggregation functions that can be used later. Alongside standard aggregation functions such as MIN and MAX, SUM, AVG, and COUNT, other functions can also be helpful, such as MEAN and STANDARD DEVIATION.

All users must implement the aggregation operation to be supported in the translator. The API includes only three functions, summarized in Table~\ref{tab:table1}. In this table, the writable data type is supported by MapReduce as an interface that stores intermediate results. The Value data type is a customizable class supported by the proposed translator, with three data components: aggregate, count, and extendable field. The aggregate part is used to aggregate the required data values, the count part stores the number of data records in the aggregation summary, and the extendable field provides an extension for user-defined aggregation.
The aggregation query consists of three steps: local aggregation, which aggregates the data elements in the map function and produces an intermediate aggregation summary; global aggregation, which aggregates the intermediate aggregation results in the aggregate function and generates a global intermediate aggregation summary; and aggregation result rendering, which translates the global intermediate aggregation summary into the final result in the desired output form.

\begin{table} 
\tbl{Descriptions of the API for Supporting User-Defined Aggregation Function.}{
\centering
\scriptsize
\begin{tabular} {| l |}
\hline
$void\ updateInMap\ (Writable,\ Value)$ \\
\ ~~Updates the local intermediate aggregation summary in the map function \\
\ ~~The arguments are the local intermediate aggregation summary and the value \\of the array element \\
\hline
$void\ updateInReduce\ (Writable,\ Writable)$ \\
\ ~~Updates the global intermediate aggregation summary in the reduce function \\
\ ~~The arguments are the local intermediate aggregation summary and the global\\ intermediate aggregation result \\
\hline
$Value\ getAggResult\ (Writable)$ \\
\ ~~Transforms the global intermediate aggregation summary into the final \\aggregation result \\
\hline
\end{tabular}
}
\label{tab:table1}
\end{table}

To show how to support non-existent user-defined aggregation functions, we assume MEAN() and STANDARD DEVIATION() as two examples and compare them with the classic AVG aggregate operator. The implementation of the API is described in Table~\ref{example2}. Herein, we describe both operations in detail. The AVG() operator should be supported by two intermediate aggregation values SUM() and COUNT(). To calculate the standard deviation, the extendable field data part is used to record the squared sum of the given array. Finally, a single transformation is applied after calculating the average. To estimate the geometric mean, the map function is used to compute the local aggregation and the log sum instead of the sum, and the transformation in the last step will be different.

To support different aggregate operators and aggregation functions, only the three functions in the translator API need to be customized. Based on this, we expect that any user-defined function can be implemented similarly. 

\begin{table*} [ht]
\tbl{Examples for Aggregation API Implementation.}{
\scriptsize
\centering
\def\arraystretch{1.5}
\begin{tabular} {| c | c | c |}
\hline
{\bf API} & {\bf STDDEV} & {\bf MEAN} \\ [0.5ex]
\hline
$updateInMap$ &  $locAgg = sum(Array);$ & $locAgg = lg\_sum(Array);$ \\
$           $ &  $locCnt = count(Array);$ & $locCnt = count(Array);$ \\
$           $ &  $locExAgg = squared\_sum(Array);$ & $          $\\
\hline
$updateInReduce$ &  $glbAgg = \sum{locAgg};$ & $glbAgg = \sum{locAgg};$ \\
$              $ &  $glbCnt = \sum{locCnt};$ & $glbCnt = \sum{locCnt};$ \\
$              $ &  $glbExAgg = \sum{locExAgg};$ & $                 $ \\
\hline
$getAggResult$ &  $avg = \frac{glbAgg}{glbCnt};$ & $lg\_avg = \frac{glbAgg}{glbCnt};$ \\
$            $ & $agg = sqrt(glbExAgg^2 - 2 \times  $ & \\
$            $ &  $glbAgg \times avg + avg^2);$ & $agg = e^{lg\_avg};$  \\
\hline
\end{tabular}
}
\label{example2}
\end{table*}

\section{Performance Optimization}
\label{opt}

In this section, we aim to optimize aggregation performance by validating two questions: Does the proposed system support efficient multidimensional array subsetting by loading just the required array subset instead of the whole array? Do the generated MapReduce jobs optimize the required aggregations and reduce the volume of intermediate results? In the following subsection, we will answer these questions and provide a detailed evaluation of the proposed system.

\subsection{Multidimensional Array Subsetting Evaluation}

Herein, we assume that no indexing mechanism has been used in our example in this section. To fulfill an SQL WHERE clause, a full scan of all the data is required to find the samples that can satisfy the WHERE clause. In this case, the samples that do not meet the given condition are ignored from the final results. This prediction type is named value-based predicate that can be faster if supported by the available system. This WHERE clause can be processed through our proposed translator without any relational table subsetting operation.

The proposed translator can scan the queried data only once and reduce the I/O cost by merging the corresponding value-based filtering phase with the data-loading stage during the aggregation. Moreover, the user sometimes needs to specify an array subset in a given array query. Formally, we refer to this kind of requirement as a dimension-based predicate, which is defined by array dimensional coordinates.

In this case, there is no need to load the entire array for query processing. However, it is highly desirable to support efficient array subsetting based on specified dimensional indices and significantly reduce the I/O cost. Unfortunately, the original Hadoop cannot support such array subsetting. Therefore, we have developed a splitter to support loading only an array subset instead of the entire input array when a dimension-based predicate exists.

\subsection{In-Mapper Aggregation}

In the context of data-intensive applications on software like Hadoop~\cite{lam2010hadoop}, significant overheads are incurred by the massive amount of intermediate results that have been produced. In addition, these results require more overhead when disk I/O operations are performed on them. Thus, many studies have tried to reduce the number of these temporary results before they are added to the disk. One way to achieve this is to use in-mapper aggregations, i.e., local aggregation/combination in the map function, before any intermediate results are written to the disk. This idea is familiar since it was already implemented in Spark~\cite{zaharia2010spark} and MATE~\cite{jiang2010map} and showed an impressive speedup.
Herein, we assume hand-written MapReduce code scripts correspond to a straightforward aggregation implementation, which is easy to write but shifts all the aggregation workload to the reducer. Moreover, the translator generates optimized code that involves local aggregation in the mapper and global aggregation in the reducer. 
First, we will take grid aggregation as an example to show the optimized overlapping aggregation algorithms with slight modifications.

\subsubsection{Naive Aggregation Implementation}

Algorithms~\ref{naive_map_grid} and~\ref{naive_reduce_grid} describe the naive map and reduce implementations. In the map phase, a key-value pair is produced for each array element in the mapper, where the key is the grid ID and the array element value. Second, a grouping operation is performed on the grid in the shuffling phase. Finally, the reducer aggregates the elements of the same group in the reduce phase. The shuffling is performed in the naive aggregation algorithm, followed by the reducer-side aggregation, where the number of intermediate results equals the number of queried elements, which can incur a considerable shuffling cost for massive datasets. Similarly, for overlapping aggregations such as sliding aggregation, the number of intermediate results equals the number of queried elements times the number of elements covered in a grid, leading to an even more costly shuffling operation.

\begin{algorithm}[htp]
\def\arraystretch{1}
\caption{NaiveMap (array split $S$)}
\label{naive_map_grid}
\begin{algorithmic}[1]
\FOR {Each array element of the value $V_i$ in \\the array split $S$}
    \STATE Identify the grid labeled as $G_j$ to which\\ the element belongs.
    \STATE Emit the key-value pair ($G_j$, $V_i$).
\ENDFOR
\end{algorithmic}
\end{algorithm}

\begin{algorithm}[htp]
\def\arraystretch{1}
\caption{NaiveReduce (grid $G_i$, array element values $V$)}
\label{naive_reduce_grid}
\begin{algorithmic}[1]
\STATE Let $A_i$ be the aggregation summary corresponding to the grid $G_i$
\STATE $A_i \leftarrow aggregate(V)$.
\STATE Output the key-value pair ($G_i$, $A_i$).
\end{algorithmic}
\end{algorithm}

\subsubsection{In-Mapper Aggregation Implementation}
Recalling Section~\ref{overview}, the proposed translator involves two types of aggregation local and global aggregations while generating the MapReduce code. The mapper manages the local aggregation while the reducer manages the global aggregation. Algorithms~\ref{opt_map_grid} and~\ref{opt_reduce_grid} show how the mapper emits local aggregation results and not all the associated array element values for each grid. In the reduction phase, all local aggregations for the same group are aggregated again to produce a global aggregation summary. The map function implementation is the major difference between optimized overlapping aggregations and grid aggregation, where each array element belonging to more than one group needs to be identified. Thus, we added another inner loop in Algorithm~\ref{opt_map_grid}.

\begin{algorithm}[htp]
\def\arraystretch{1}
\caption{OptimizedMap (array split $S$)}
\label{opt_map_grid}
\begin{algorithmic}[1]
\STATE Initialize all the local aggregation summary $A$, where $A_i$ corresponds to the grid $G_i$.
\COMMENT {local aggregation}
\FOR {Each array element of the value $V_i$ in \\ the array split $S$}
    \STATE Identify the grid $G_j$ that the element belongs to.
    \STATE $A_j \leftarrow aggregate(A_j, V_i)$.
\ENDFOR
\FOR {Each local aggregation result $A_i$ in $A$}
    \STATE Emit the key-value pair ($G_i$, $A_i$).
\ENDFOR
\end{algorithmic}
\end{algorithm}

\begin{algorithm}[htp]
\def\arraystretch{1.5}
\caption{OptimizedReduce(grid $G_i$, local aggregation results $A^{loc}$)}
\label{opt_reduce_grid}
\begin{algorithmic}[1]
\STATE Let $A^{glb}_i$ be the aggregation summary corresponding to the grid $G_i$.
\STATE $A^{glb}_i \leftarrow aggregate(A^{loc})$.
\STATE Output the key-value pair ($G_i$, $A^{glb}_i$).
\end{algorithmic}
\end{algorithm}

In Algorithm~\ref{opt_map_grid}, an accumulation operation in the map function is used to collect the local aggregation results. This kind of operation only works for algebraic aggregate operators, i.e., holistic aggregate operators like MEDIAN() and RANK() are not suitable for this type. In conclusion, the in-mapper aggregation can add massive speedup by reducing the volume of intermediate results written to the disk and shuffled across the network.

\begin{table*}[!ht]

\newcommand{\tabincell}[2]{\begin{tabular}{@{}#1@{}}#2\end{tabular}}
\tbl{Example AQL Queries.}{
\def\arraystretch{1.5}
\scriptsize
\centering
\begin{tabular}{|l|l|}
\hline
\multicolumn{2}{|l|}{\bf Example Grid Aggregation Queries:}\\
\hline
Full Grid Aggregation & select avg(Val) from L1 grid as (partition by x 512 y 512)\\
\hline
Grid Aggregation (50\% subset) & select avg(Val) from between\\
& (L2, 0, 0, 16383, 32767) grid as (partition by x 512, y 512)\\
\hline
Grid Aggregation (25\% subset) & select avg(Val) from between 
\\& (L1, 16384, 0, 24575, 32767) grid as (partition by x 512, y 512)\\
\hline
Grid Aggregation (13\% subset) & select avg(Val) from between
\\& (L2, 24576, 0, 28671, 32767) grid as (partition by x 512, y 512)\\
\hline
\multicolumn{2}{|l|}{\bf Example Sliding Aggregation Queries:}\\
\hline
Full Sliding Aggregation & select avg(Val) from L1 fixed window as (partition by x 1 preceding and \\& 1 following, y 1 preceding and 1 following)\\
\hline
Sliding Aggregation (50\% subset) & \tabincell{c} {select avg(Val) from between (L2, 0, 0, 4999, 9999) fixed window as \\(partition by x 1 preceding and 1 following, y 1 preceding and 1 following)}\\
\hline
Sliding Aggregation (25\% subset) & \tabincell{c} {select avg(Val) from between (L1, 5000, 0, 7499, 9999) fixed window as \\(partition by x 1 preceding and 1 following, y 1 preceding and 1 following)}\\
\hline
Sliding Aggregation (13\% subset) & \tabincell{c} {select avg(Val) from between (L2, 7500, 0, 8749, 9999) fixed window as \\(partition by x 1 preceding and 1 following, y 1 preceding and 1 following)}\\
\hline
\end{tabular}
}
\label{table:queries}
\end{table*}

\section{Experimental Results}
\label{expr}

In this section, we evaluate the performance of the proposed translator from different aspects. First, the evaluation involves several query types, including grid and sliding aggregations, over the whole data array or its subset. Second, the assessment also involves user-defined aggregation functions such as MEAN() and STANDARD DEVIATION(). Third, we also aim to assess the benefits of utilizing the optimizations that are supported by the proposed system, such as the subset splitting of the input and in-mapper aggregation.

The experiments were conducted on four computing nodes, each comprising 16 Intel Xeon Intel E5640 cores with 2.67GHz. The main memory on each node is 24GB. The proposed system was built on top of Hadoop-1.2.1. The dataset used for grid aggregations is a 4GB $32{,}768\times32{,}768$ floating-point 2D data array. We used a $10{,}000\times10{,}000$ floating-point 2D data array for evaluating sliding aggregations.

\subsection{Evaluating Array Subsetting Optimization}

Table~\ref{table:queries} shows AQL query examples assessed on our proposed system. To evaluate the performance of the grid and sliding aggregation, we use the entire array and its subsets, ranging from 50\% to 12.5\%, with the goal of subsetting predicates over all the dimension-based predicates. The evaluations of both aggregations include the version with and without array subsetting support to show the benefits of utilizing array subsetting at input splitting time.

\subsubsection{Evaluating Grid Aggregation}

\begin{figure}[htbp]
    \centering
    \includegraphics[width=0.8\linewidth]{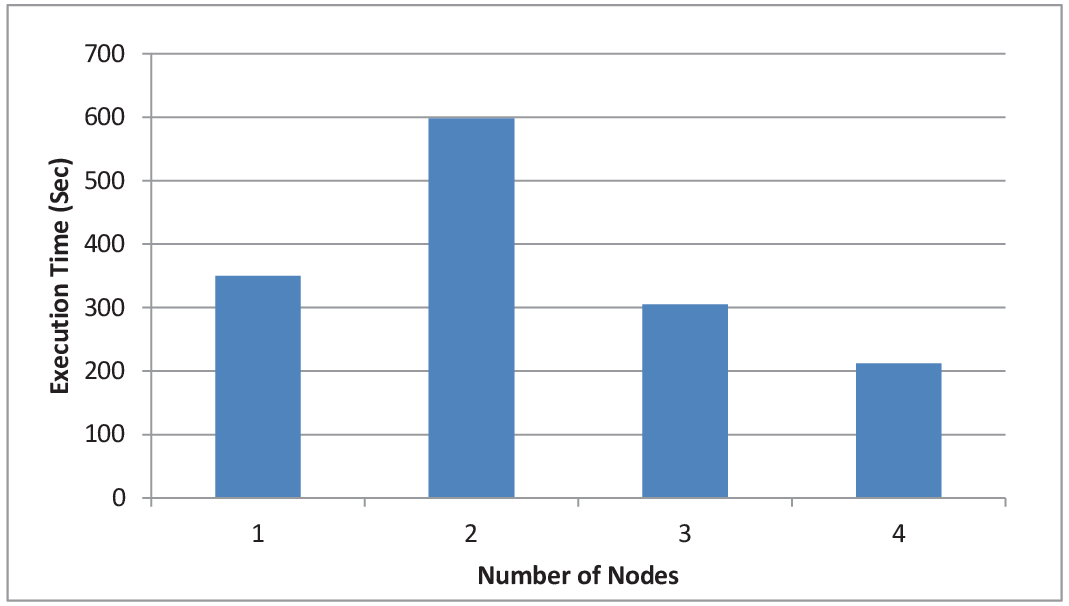}
    \caption{Full Grid aggregation execution time.}
    \label{grid}
\end{figure}
    \hfill
\begin{figure}[htbp]
   \centering
    \includegraphics[width=0.8\linewidth]{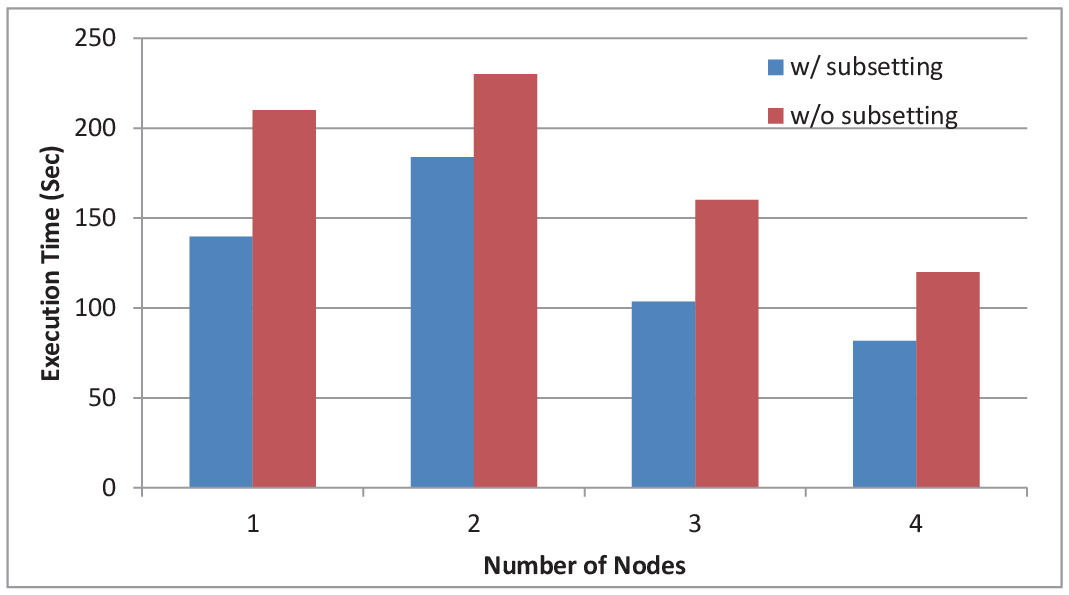}
    \caption{50\% Subset Grid aggregation execution time.}
    \label{grid50}
\end{figure}
 
\begin{figure}[htbp]
   \centering
    \includegraphics[width=0.8\linewidth]{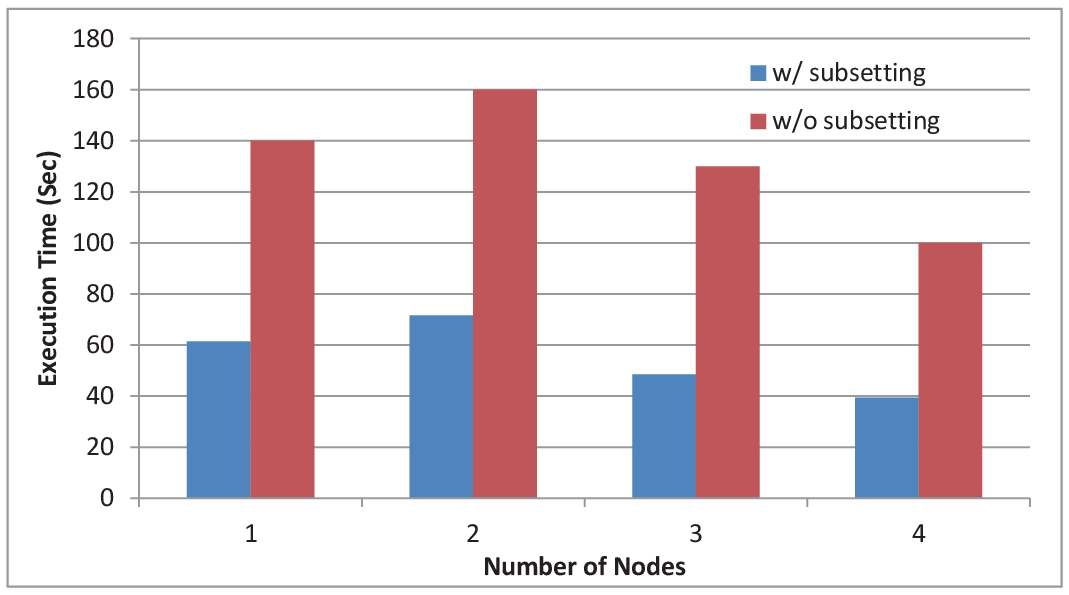}
    \caption{25\% Subset Grid aggregation execution time.}
    \label{grid25}
\end{figure}

\begin{figure}[htbp]
   \centering
    \includegraphics[width=0.8\linewidth]{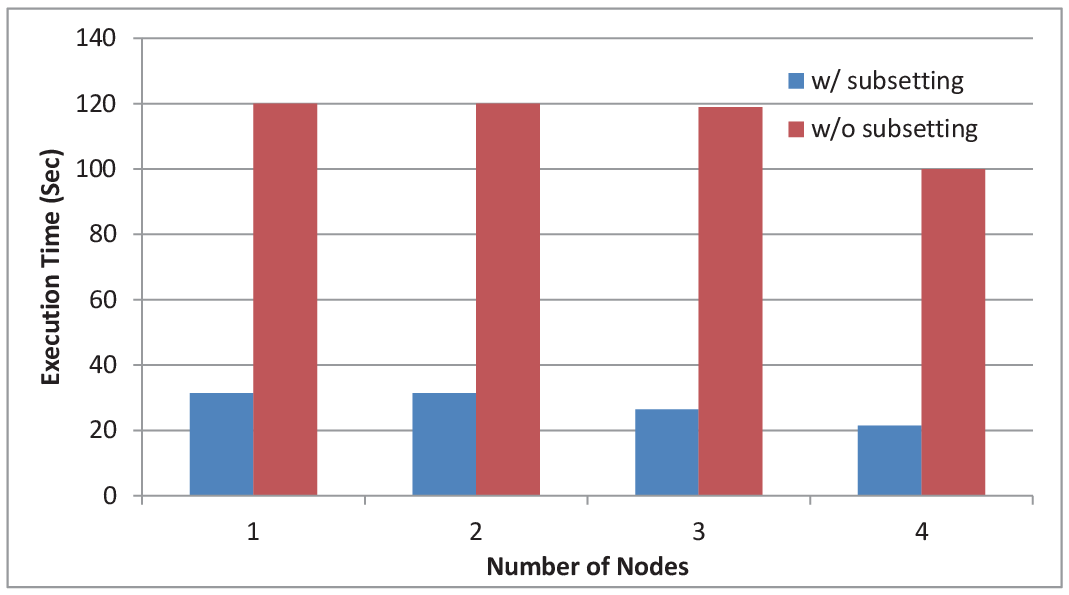}
    \caption{12.5\% Subset Grid aggregation execution time.}
    \label{grid13}
\end{figure}

Figure~\ref{grid} illustrates the execution time for the entire grid aggregation with varying numbers of nodes: 1, 2, 3, and 4. The figure reveals that increasing the number of nodes from one to two results in a $1.71$X slowdown in query evaluation. This behavior is attributed to two main factors: the overhead of remote data access and the scheduling overhead required to coordinate additional computing nodes. These overheads do not increase linearly with the number of nodes. Consequently, the speedup achieved with four-node execution is $1.65$X. Therefore, increasing the number of nodes helps mitigate the impact of remote data access and scheduling overheads.

The execution time for grid aggregations with and without subsetting is shown from Figure~\ref{grid50} to Figure~\ref{grid13}. The performance is similar for all grid aggregation queries for the versions with subset splitting support. For example, the speedups of four-node executions are $1.71$X, $1.56$X, and $1.47$X for 50\%, 25\%, and 13\% queries, respectively. Compared with the versions without array subsetting support, the version with subsetting support for 50\% subset query achieves a $1.41$X speedup. For a 25\% subset query, a high speedup of $2.4$X is achieved as the array subsetting support avoids even more unnecessary data loading. In contrast, the version without subsetting must still load all the input. For a 13\% subset query, the speedup is $4.15$X.

\subsubsection{Evaluating Sliding Aggregation}

\begin{figure}[hp]
    \centering
    \includegraphics[width=0.8\linewidth]{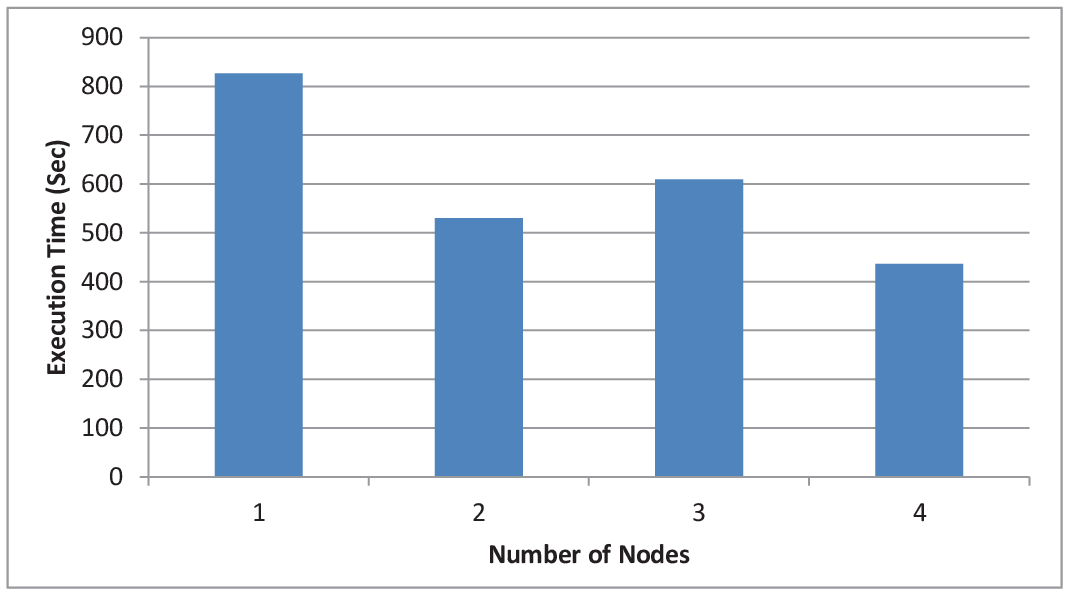}
    \caption{Full sliding aggregation execution time.}
    \label{slide}
\end{figure}
    \hfill
\begin{figure}[hp]
\centering
    \includegraphics[width=0.8\linewidth]{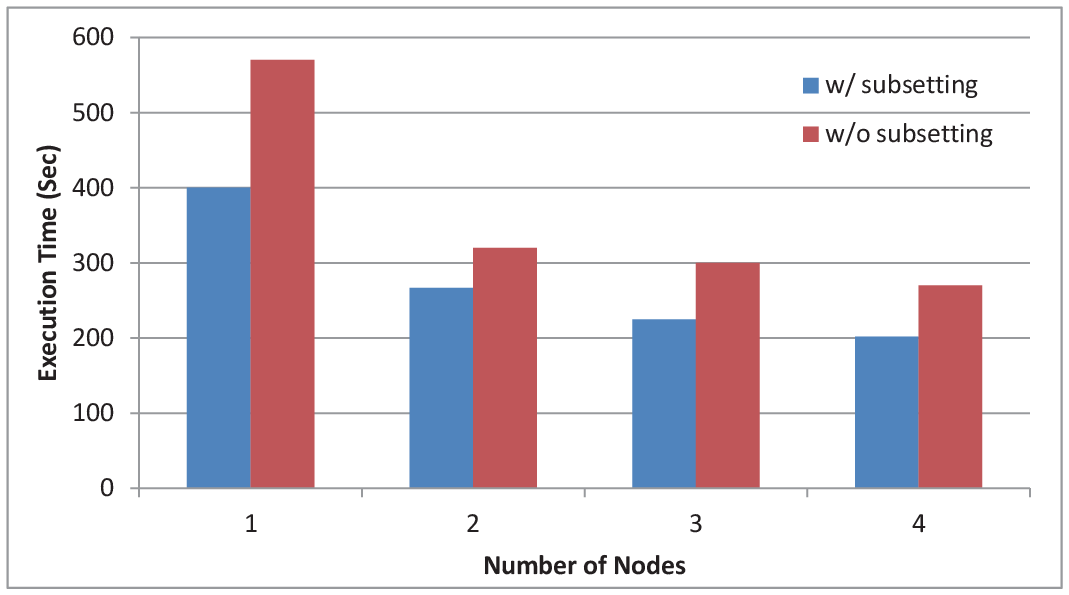}
    \caption{50\% subset sliding aggregation execution time.}
    \label{slide50}
\end{figure}
     \\
\begin{figure}[hp]
\centering
    \includegraphics[width=0.8\linewidth]{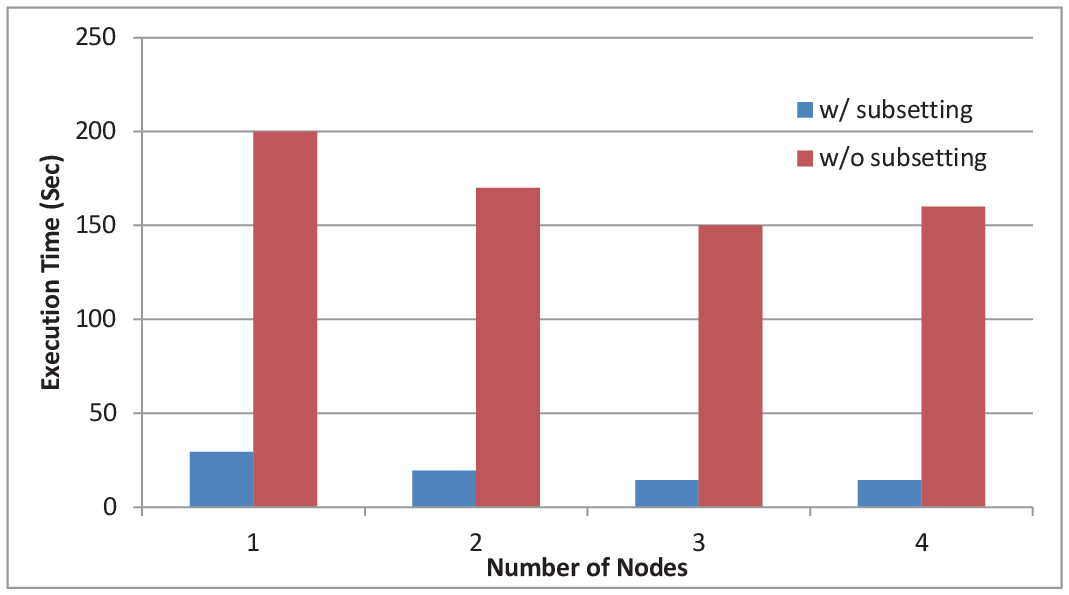}
    \caption{25\% subset sliding aggregation execution time.}
    \label{slide25}
\end{figure}
     \hfill
\begin{figure}[hp]
\centering
    \includegraphics[width=0.8\linewidth]{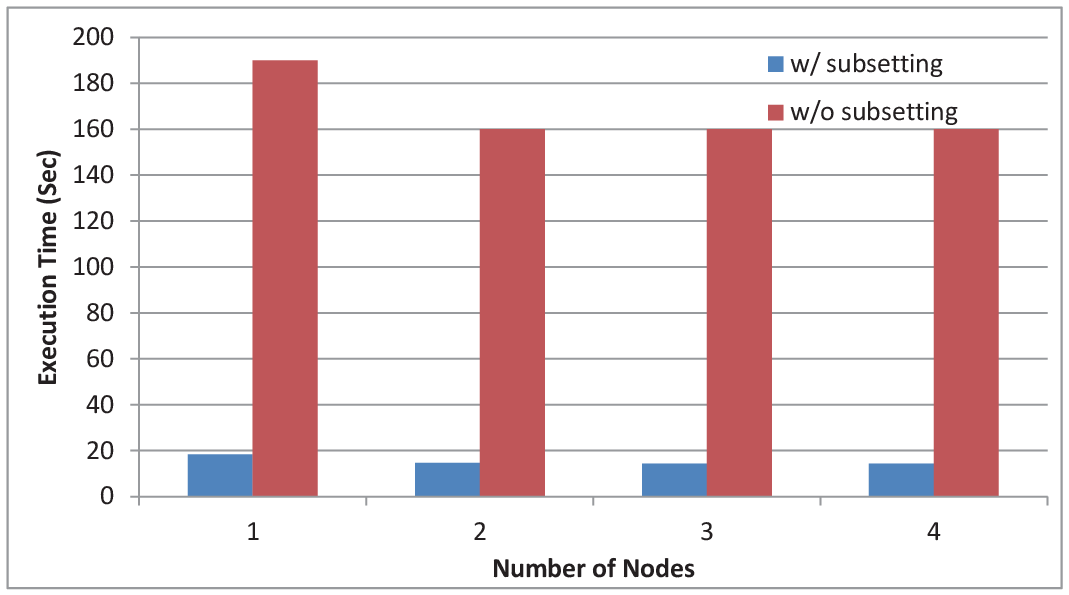}
    \caption{12.5\% Subset sliding aggregation execution time.}
    \label{slide13}
\end{figure}

Figure~\ref{slide} shows the execution time for the sliding aggregation on the whole input dataset. Similar to the grid aggregation query, when the number of nodes increases from two to three, the performance drops by $1.15$X; this is mainly caused by data access locality, though four nodes still achieve a speedup of $1.90$X over using one node only. Figures~\ref{slide50} to~\ref{slide13} show the execution time for the sliding aggregation with different subsetting ratios. Overall, the execution time decreases with the number of computing nodes increasing for every query processing. The speedups of four nodes are $1.98$X, $2.04$X, and $1.28$X for each subsetting ratio. Notice that when the subsetting ratio decreases from 50\% to 25\%, the execution time decreases by $13.79$, which is super-linear. The reason is that when the amount of data to be queried is decreased, the access locality is significantly increased. Again, the array subsetting support delivers speedups to the three subset queries of $1.33$X, $8.76$X, and $10.84$X, respectively, showing the efficiency of our array subsetting support.

\subsubsection{Evaluating Subsetting with Combined Subsetting Predicates}

In the previous subsection, we showed the results of subsetting with dimension-based predicates. We assess the subsetting with value-based predicates and dimension-based predicates, i.e., the combined subsetting predicates. The proposed translator performs the value-based filtering in the mapper instead of the splitter. Because such value-based filtering is merged with the mapping phase, the extra execution cost is relatively small compared with the original computation costs. Therefore, the results are trivially different from the ones obtained in the prior experiments, and the reader can draw the same conclusion. The only difference in the AQL queries is adding the WHERE clause. For example, 'WHERE $age > 33$ indicates only the array elements of the values greater than $33$ are involved in the aggregation.

\subsection{Evaluating In-Mapper Aggregation Optimization}

In this subsection, we compare the performance of MapReduce-generated code by the proposed translator with the hand-written MapReduce code to assess the effectiveness of the generated code. We rely on up to four nodes to perform our experiments, and the entire array is being processed.

\begin{figure}[htbp]
\centering
    \includegraphics[width=0.8\linewidth]{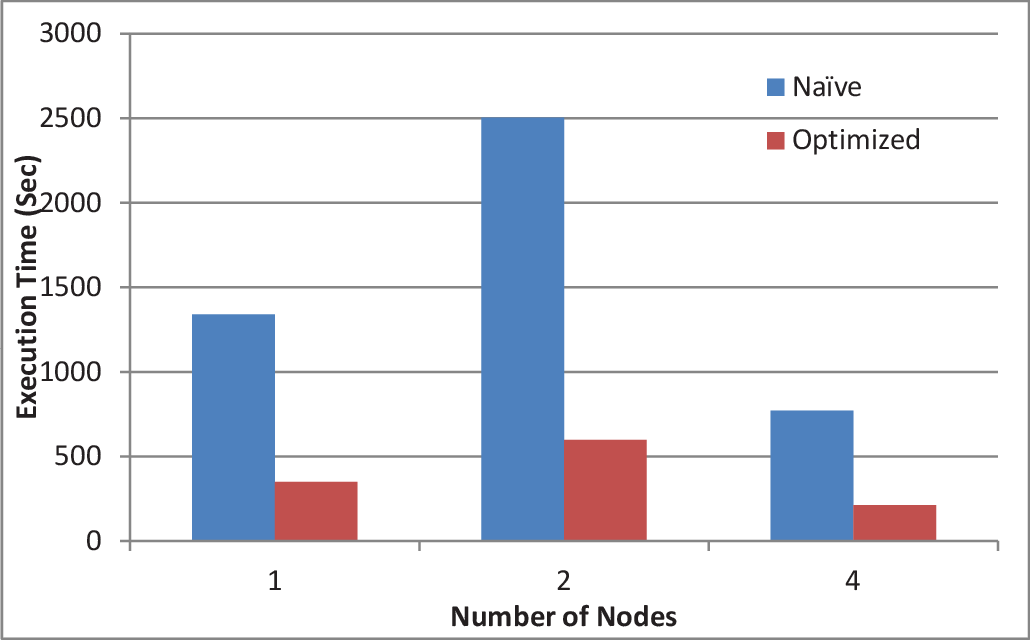}
    \label{optgrid1}
    \caption{Grid aggregation optimization evaluation.}
    \label{opt_eval1}
\end{figure}

\begin{figure}[htbp]
\centering
    \includegraphics[width=0.8\linewidth]{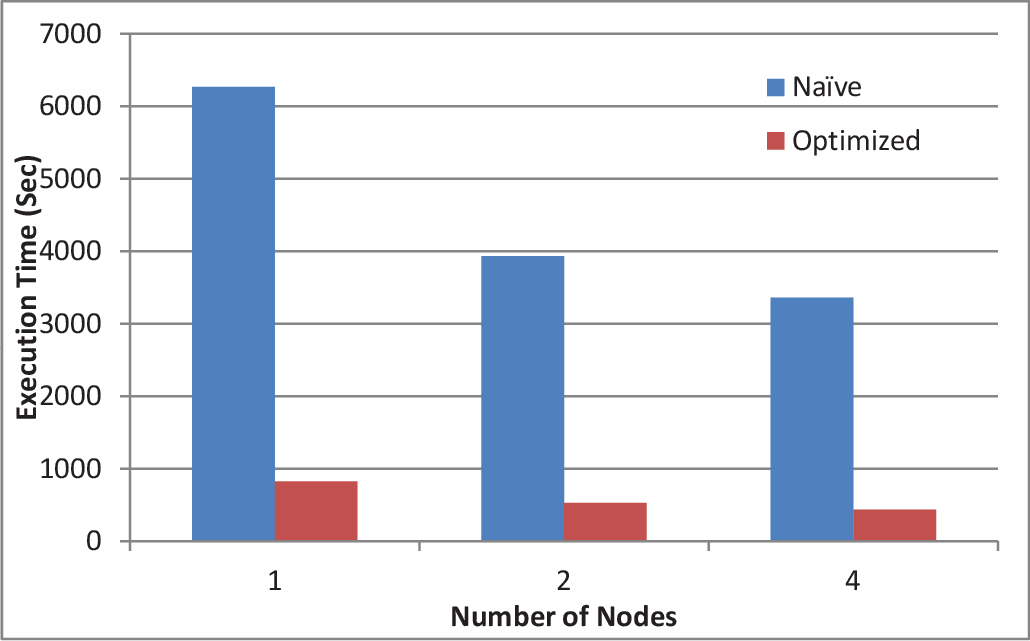}
    \label{optslide2}
    \caption{Sliding aggregations optimization evaluation.}
    \label{opt_eval2}
\end{figure}

Figures~\ref{opt_eval1} and~\ref{opt_eval2} show the grid and sliding aggregations results, respectively. It is obvious from the two figures that the performance of the generated MapReduce code is better than that of the hand-written code. The performance gain comes from the optimization step in the proposed translator, which involves reducing the massive volumes of intermediate results through the mappers to reduce the shuffling costs. Moreover, it can clearly be shown that the sliding aggregation result shows a higher performance when it comes from our translator. This is because, without the in-mapper aggregation, vast volumes of intermediate results are generated, making the execution extremely slow.

In summary, the experiments show how effective our proposed translator is, which can help perform AQL to MapReduce translation with better performance and less effort than the hand-written code. Moreover, the results show that the performance gains increase with the number of used nodes by removing all the intermediate results that can be generated through the aggregation process and allowing direct result collections.

\section{Related Work}
\label{related}

In the literature, numerous efforts have been made to improve the array database systems, the array query languages, and the translators to bridge the relational databases into MapReduce jobs. For instance, in ~\cite{Brown10}, SciDB, an array DBMS related to multidimensional data arrays, has been proposed. SciDB provides a translator to translate the SQL queries to MapReduce code but with less efficient performance than our proposed translator. Moreover, it only covers some of the aggregation operations mentioned in this paper. We observed better performance in applying the structural aggregations through our translator by eliminating expensive data ingestion steps needed by SciDB.
Moreover, in~\cite{Baumann98}, RasDaMan DBMS has been proposed, considered an algebraic-based system. It is commonly used abundantly to support different scientific applications~\cite{Baumann99}. Examples of DBMS include ArrayDB~\cite{marathe2002query}, an array database system primarily used to process small two-dimensional images; MonetDB~\cite{MonetDB}, a column-store database management system for spatial/spatio-temporal applications. RasDaMan, MonetDB, and SciDB all support exact structural aggregations similar to the proposed system. All these systems have studied different query languages and operators. For instance, SciDB~\cite{Brown10} supports both an SQL-like query language AQL and a functional language Array Functional Language (AFL); RasDaMan uses RasQL~\cite{Baumann98}, and MonetDB initially used both RAMand SciQL~\cite{IDEAS11}. More recent systems are Chronosdb~\cite{zalipynis2018chronosdb} and ArrayBridge~\cite{xing2018arraybridge}.

On the other hand, as a popular programming model, MapReduce has also been leveraged to facilitate query processing and used as a database engine~\cite{thusoo2009hive,lee2011ysmart, pang2021aqua+}. Several translators, including Pig Latin/Pig~\cite{olston2008pig}, SCOPE~\cite{chaiken2008scope}, HiveQL~\cite{thusoo2009hive}, YSmart~\cite{lee2011ysmart}, and Gerenuk~\cite{navasca2019gerenuk}, have been developed to allow the queries expressed in SQL-like query languages to be translated into MapReduce scripts automatically. However, to the best of our knowledge, there is no effort to translate any array query language into MapReduce code.

\section{Conclusions}
\label{cons}

This paper has presented both the design and the implementation of a translator that supports translating an array query language (AQL) into MapReduce code. Specifically, we focus on helping three types of structural aggregation that appear in several scientific applications. We have demonstrated that, by the elegant system design and the translation with nuanced algorithms, efficient and scalable structural aggregations, even with user-defined aggregation functions, can be effectively supported by MapReduce without extra programming effort. Furthermore, we have shown that our generated MapReduce code can lead to significantly better performance than straightforward hand-coded MapReduce scripts. Finally, we have demonstrated that our generated MapReduce scripts can efficiently process array subsets and scale well in distributed environments with customized array processing modules. 

\bibliographystyle{ijcaArticle}
\bibliography{ref}

\end{document}